\documentclass[letterpaper,prl,twocolumn,showpacs,amsmath,amssymb]{revtex4}

\usepackage{graphicx}
\usepackage{dcolumn}
\usepackage{bm}
\usepackage{soul}

\begin{document}


\title{Experimental Demonstration of Time-Delay Interferometry\\ for the Laser Interferometer Space Antenna}

\author{Glenn de Vine}
\email{glenn.devine@jpl.nasa.gov}
\author{Brent Ware}%
\author{Kirk McKenzie}%
\author{Robert E. Spero}%
\author{William M. Klipstein}%
\author{Daniel A. Shaddock}
 \altaffiliation[Also at ]{Department of Physics, The Australian National University, Canberra, Australia.}

\affiliation{Jet Propulsion Laboratory, California Institute of Technology, Pasadena, CA}

\date{\today}

\begin{abstract}
We report on the first demonstration of time-delay interferometry (TDI) for LISA, the Laser Interferometer Space Antenna. TDI was implemented in a laboratory experiment designed to mimic the noise couplings that will occur in LISA. TDI suppressed laser frequency noise by approximately $10^9$ and clock phase noise by $6\times10^4$, recovering the intrinsic displacement noise floor of our laboratory test bed. This removal of laser frequency noise and clock phase noise in post-processing marks the first experimental validation of the LISA measurement scheme.
\end{abstract}

\maketitle

%
The Laser Interferometer Space Antenna (LISA)\,\cite{Bender} is a joint National Aeronautics and Space Administration (NASA) and 
European Space Agency (ESA) gravitational wave observatory. LISA will observe gravitational radiation from: massive black hole 
mergers out to a distance of z=20; black hole, neutron star, and white dwarf inspirals into massive black holes; and white-dwarf binary orbits 
throughout the galaxy\,\cite{LISASC2007}.

LISA will measure the relative motion of three drag-free spacecraft (SC) separated by $5\times10^{9}$\,m with a one-way resolution of $2 \times 10^{-11}$\,m/$\sqrt{\rm Hz}$ ($4\times10^{-21}/\sqrt{\rm Hz}$ strain sensitivity). Laser light is passed between SC and the interference phase between the local and distant laser (one-way) phases recorded. The design sensitivity is dominated by shot noise from the laser light for frequencies above 3\,mHz and by spurious forces on the proof masses at lower frequencies.

Orbital motion of the SC Doppler shifts the laser beams by up to 20\,MHz, giving rise to heterodyne signals upon interference with a local oscillator on each SC. The phase change of these beat note signals is proportional to the change in path length between SC. Gravitational waves also cause a displacement between SC, phase-shifting the beat note. The challenge for LISA is to measure these phase shifts with $\mu$cycle accuracy in the presence of slowly varying Doppler shifts, millions of cycles of laser frequency noise and variations in the clock sampling frequencies.

The LISA arm lengths will neither be matched (the mismatch will be up to 75,000\,km) nor static, introducing sensitivity to laser frequency noise. LISA will use a technique called time-delay interferometry (TDI), combining local and inter-spacecraft phase measurements in post-processing, to form configurations equivalent to Michelson and Sagnac interferometers\,\cite{Tinto99}. TDI suppresses noise from laser frequency fluctuations by many orders of magnitude, yet preserves the gravitational wave signal. TDI consists of linear combinations of the phase measurements recorded at specific times determined by the light travel time between SC. TDI will also correct for phase noise of the ultra-stable oscillators (or clocks), which provide the phase measurement references on each SC. 

Although TDI has been extensively studied theoretically, there have not previously been any experimental demonstrations of the key aspects of the signal processing chain. This paper reports results from the first demonstration of TDI in a laboratory experiment designed to mimic the noise couplings that will occur in LISA. This experiment was set up to resemble two LISA SC, each with two lasers and a phasemeter referenced to an independent clock. The results show that TDI suppressed laser frequency fluctuations by $10^9$ at 3\,mHz and clock noise by $6\times10^4$. This confirms the LISA measurement scheme and validates the performance of the LISA phasemeter\,\cite{LISApm} with a LISA-like signal structure.  

TDI can be understood as a technique to synthesize equal arm-length interferometer configurations from one-way measurements. With equal lengths, the effect of laser frequency noise is common to each arm of a two beam interferometer and will cancel when differenced. 

\begin{figure}[ht]
\includegraphics[width=8.8cm]{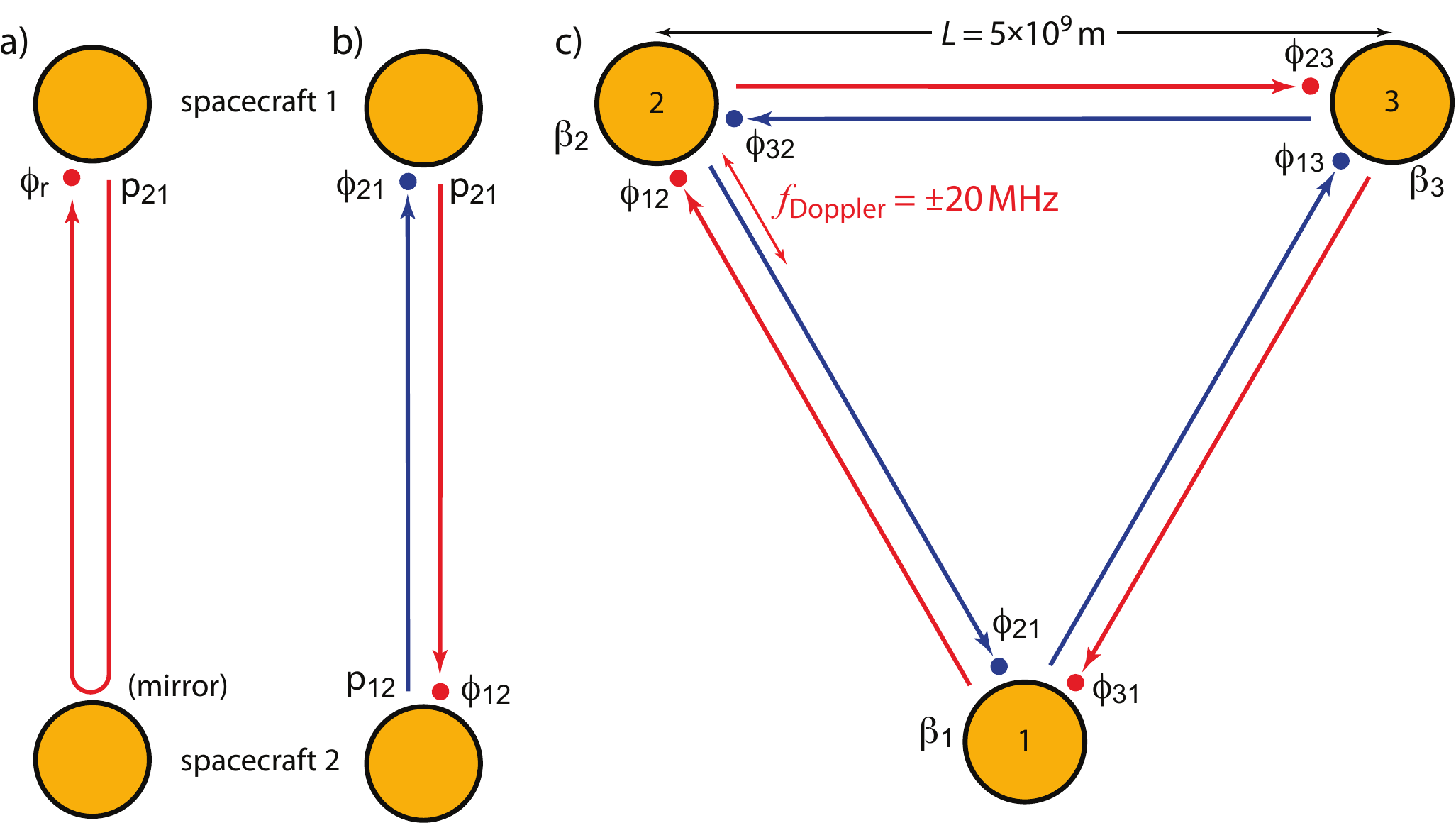}
\caption{Schematic representation of laser links between spacecraft. a) single round-trip and b) one-way phase measurements. c) LISA constellation showing the six (one-way) inter-spacecraft phase measurements. All spacecraft are identical.}
\label{fig:alpha} 
\end{figure}

To illustrate the concept of a synthetic configuration, we show how a round-trip measurement, Fig.\,\ref{fig:alpha}a, can be made by combining two one-way measurements, 
Fig.\,\ref{fig:alpha}b. In Fig.\,\ref{fig:alpha}a a laser beam travels a distance $L_{12}$ from SC1 to SC2, where it is retro-reflected, traveling a further distance, $L_{21}$. The phase of the return beam is measured relative to the outgoing beam, $\phi_r(t) =D_{21}D_{12}p_{21}(t)- p_{21}(t) $, where $p_{ij}(t)$ is the output laser phase on SC $j$ looking at SC $i$. $D_{ij}$ is a delay operator\,\cite{Dhurandhar} representing the application of a time delay: $D_{ij}a(t) = a(t-L_{ij}/c)$. In this paper we deal only with static arm lengths, therefore delay operators commute, $[D_{i},D_{j}] = 0$. For clarity, noise sources other than laser phase noise are neglected, until Eq.\,\ref{eq:AlphaClocks}. The same information in $\phi_r(t)$ can be acquired by making two one-way measurements (Fig.\,\ref{fig:alpha}b) and combining them with a time delay determined by the light travel time. 
LISA employs this symmetric arrangement, with a laser and a phase measurement made at each end. The measurements at SC1 and 2, respectively, are $\phi_{21}(t) =D_{21}p_{12}(t)- p_{21}(t)$ and $\phi_{12}(t) =D_{12}p_{21}(t)- p_{12}(t)$\,\footnote{Note that we have assumed simultaneity of the phase measurements taken on the separate spacecraft.}. Combining these one-way links with a time-shift gives the same result as the conventional mirror-based measurement: $\phi_{21}(t)+D_{21}\phi_{12}(t) = \phi_r(t)$.

Extending this approach to multiple links allows numerous interferometer configurations to be formed, including combinations that maintain equal arm lengths in the presence of SC velocity \cite{Shaddock03,TintoPRD04}. A conventional Sagnac interferometer would be formed by interfering two laser beams that have counter-propagated around the constellation. The same information is contained in the TDI combination, $\alpha$\,\cite{Shaddock042}:
\begin{eqnarray}
\alpha(t) = \phi_{31}(t)+D_{31}(\phi_{23}(t)-\beta_3(t))+D_{23}D_{31}(\phi_{12}(t)-\beta_2(t))\nonumber \\-\phi_{21}(t)-D_{21}
(\phi_{32}(t)+\beta_2(t))-D_{32}D_{21}(\phi_{13}(t)+\beta_3(t))\nonumber \\-(1+D_{12}D_{23}D_{31})\beta_{1}(t), \nonumber \\
\label{eq:AlphaFig}
\end{eqnarray}
In Eq.\,\ref{eq:AlphaFig} we have included the back-link measurements, $\beta_j(t)$, made on SC $j$\,; a measure of the 
phase difference between the two local lasers: $\beta_1(t) = p_{21}(t)-p_{31}(t)$; $\beta_2 (t)= p_{32}(t)-p_{12}(t)$; and $\beta_3(t) = p_{13}(t)-p_{23}(t)$. The phases of the inter-spacecraft links from SC $i$ to SC $j$ are $\phi_{ij}(t) = D_{ij}p_{ji}(t)-p_{ij}(t)$.

The phase measurements will be triggered by an onboard clock. As such the phase measurements taken on separate SC will not be taken simultaneously (this effect is not included in Eq.\,\ref{eq:AlphaFig}). To adequately suppress the frequency noise, described by Eq.\,\ref{eq:AlphaFig}, in addition to correcting for the light travel time between spacecraft, the delay operators must correct for the relative clock offset, to the level of $\sim3$\,ns\,\cite{FCST}. LISA will use a dedicated system to determine both the light travel times and the clock offsets. The phase measurements will be sampled at $\approx3$\,Hz and later shifted to the required times using ns-accuracy interpolation algorithms\,\cite{Shaddock04}.

The spacecraft clocks also introduce clock phase noise into each phase measurement. Clock phase fluctuations are indistinguishable from beat note fluctuations, which contain the gravitational wave signal. To measure the phase of a 20\,MHz (maximum Doppler shift, $f_{0}$) beat note with a phase sensitivity, $\phi(f) = 10^{-6}\,{\rm cycles}/\sqrt{\rm Hz}$ at $f\,=\,3$\,mHz, LISA requires a clock stability of: $y(f) = 2\pi f\phi(f)/f_{0} = 9 \times 10^{-16}/\sqrt{\rm Hz}$.

With the required stability unavailable for spaceflight, the clock noise will be measured and removed. This can be achieved by transferring the clock phase between spacecraft with a fidelity of $\mu$cycles/$\sqrt{\rm Hz}$\,\cite{Hellings, Klipstein06}. To improve the signal to noise ratio, the clock signal will be multiplied to a frequency of several GHz and phase-modulated onto the laser. Interference between the distant and local lasers' sidebands appear in the photodetector signal near the carrier-carrier beat note frequency, and can be measured by the phasemeter. These sideband-sideband beat notes give a measure of the clock noise, which can be removed by modified TDI combinations in data processing. 

\begin{figure}[htb]
\includegraphics[width=8cm]{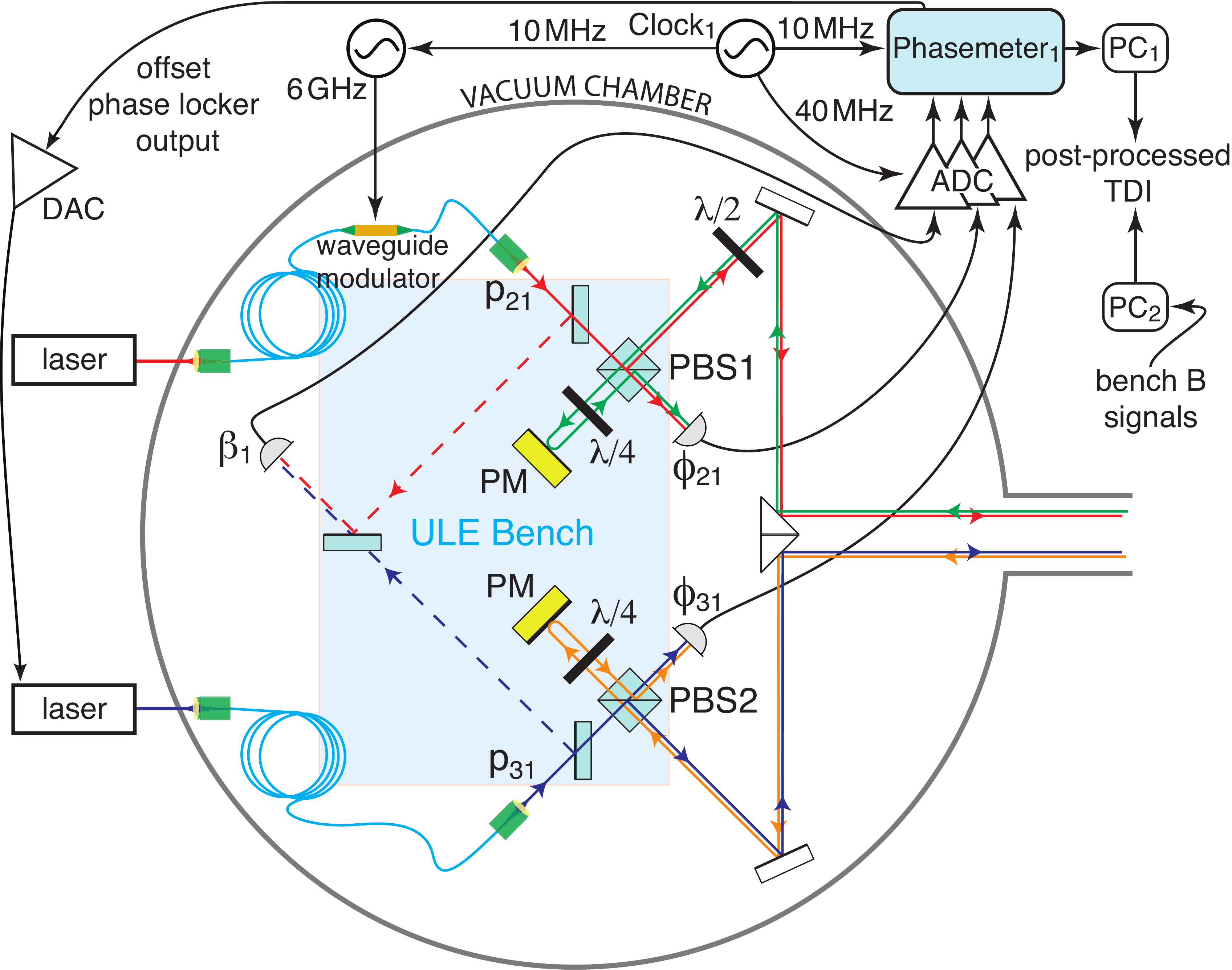}
\caption{Schematic of one bench of the LISA interferometry
  test bed (the other bench is an identical copy). The benches are mechanically coupled via a vacuum envelope. $\lambda/4$ = quarter wave-plate, $\lambda/2$ = half wave-plate, 
PBS = polarizing beam-splitter, PM = proof mass, DAC = digital-to-analog converter, ADC = analog-to-digital converter.}
\label{fig:test bed}
\end{figure}

The LISA interferometry test bed (Fig.\,\ref{fig:test bed}), based on the TDI Sagnac combination $\alpha$, is designed to provide signals representative of LISA interferometry, without requiring 5 million km arms.  The experiment reproduces the essential experimental complexity of LISA: polarization leakage interferometry; multiple heterodyne frequencies; independent clocks; and independent phase measurements of optical signals and clock sidebands. Multiple heterodyne frequencies are required to avoid artificial common-mode cancellation of non-linear effects. A representative signal structure for testing these concepts is obtained with two optical benches, avoiding unnecessary experimental complexity of three benches.

With only two benches in our experiment, the $\alpha$ combination must be modified. This is equivalent to the lasers on the third SC being phase-locked according to the scheme presented in \cite{Shaddock042}. In this case the inter-spacecraft links previously containing SC3 become $\phi_{32}(t) = D_{32} D_{13}p_{31}(t)-p_{32}(t),~~\phi_{31} (t)=D_{23}D_{31}p_{32}(t)-p_{31}(t)$, with $\beta_3(t)=\phi_{23}(t)=\phi_{13}(t)=0$. Assuming two SC with independent clocks, Eq.\,\ref{eq:AlphaFig} becomes:
\begin{eqnarray}
\alpha^{c}(t_1,t_2) = \phi_{31}^c(t_1)+D_{23}D_{31}(\phi_{12}^c(t_2)-\beta_2^c(t_2))\nonumber \\-\phi^{c}_{21}(t_1)-D_{21}
(\phi_{32}^c(t_2)+\beta_2^c(t_2))\nonumber \\-(1+D_{12}D_{23}D_{31})\beta_{1}^c(t_1), 
\label{eq:AlphaClocks}
\end{eqnarray}
where the notation change $\phi_{ij}(t)\rightarrow\phi_{ij}^c(t_j)$ indicates that each measurement is made at time $t_j$, triggered by the clock on SC $j$, and the superscript $c$ denotes a phase measurement containing clock phase noise. The normalized clock time error, $q_j(t_j)$, enters each phase measurement made on SC $j$ in direct proportion to its heterodyne beat note frequency. The single link and back-link measurements become, for example:
\begin{equation}
\begin{array}{llllll}
\phi_{ij}^c(t_j) &=&&\phi_{ij}(t_j)&-& q_j(t_j)(\nu_{ji}(t_j)-\nu_{ij}(t_j))\\
\beta_j^c (t_j) &=&&\beta_j(t_j) &-& q_j(t_j)(\nu_{ij}(t_j)-\nu_{kj}(t_j)),\\
\end{array}
\end{equation}
where $\nu_{ij}(t)$ is the optical frequency of the $ij$ laser.

The experiment comprises two simplified LISA-topology optical benches constructed from ultra-low expansion (ULE) glass (170$\times$120$\times$20\,mm), separated by 1.0\,m, and occupying a common vacuum envelope ($\sim$1\,mTorr). For sensitive optical paths, fused silica components were optically contacted onto the ULE bench. This technique has shown picometer-level displacement stability on a single bench\,\cite{Shaddock03spie}.
Each optical bench represents a single spacecraft; one is illustrated in Fig.\,\ref{fig:test bed}. The two benches each have two lasers and three phase measurements: two inter-bench, $\phi_{21}$ and $\phi_{31}$ on bench 1, and $\phi_{12}$ and $\phi_{32}$ on bench 2; and one local back link, $\beta_{1}$ on bench 1, and $\beta_{2}$ on bench 2 (all photodetectors: 125\,MHz bandwidth, 2.5\,pW/$\sqrt{\rm{Hz}}$ NEP, $\sim$1\,$\mu$W incident power). The two lasers (Nd:YAG, 500\,mW) on each bench were phase-locked (unity gain, $f_{\rm{ug}} \approx 30$\,kHz) with an offset heterodyne frequency ($ 4.25$\,MHz on Bench 1; $3.90$\,MHz on Bench 2). The laser $p_{12}$ was phase locked to $p_{21}$ (as will be done between spacecraft), offset at $ 4.00$\,MHz, mimicking a static Doppler shift. White frequency noise was added to the error point of two of the three phase-locking control loops: $\beta_1(t)\approx n_1(t)$ and $\phi_{12}(t)\approx n_2(t)$ at a level of $|n_{1}(t)|/(2\pi f)=|n_{2}(t)|/(2\pi f) =800$\,Hz/$\sqrt{\rm Hz}$, mimicking a pre-stabilized noise of that level for the LISA lasers. Because of the short path length, $\alpha$ is first-order insensitive to path length noise in the common optical lengths. Therefore, picometer stability in the arms (i.e., between benches) was not necessary. However, $\alpha$ is sensitive to optical path changes within each ULE bench, e.g. proof-mass displacement.

Several mirrors in the nominally insensitive optical path were dithered at tens of Hz with 1\,cycle amplitude triangular waveforms, in order to up-convert cyclic nonlinearity from spurious interferometer paths to frequencies above the LISA signal band. The amplitude of the cyclic nonlinearity without dither was approximately 1\,nm, and the frequency, equal to the number of wavelengths/second of thermal expansion in the Sagnac path, was in the mHz band. The LISA optical bench will also suffer from cyclic nonlinearity from spurious interference, though of reduced amplitude owing to better control of scatter. Dither on LISA will not be necessary as the intra-spacecraft optical paths, by design, should drift by much less than a wavelength.

\begin{figure}[htb]
\includegraphics[width=8cm]{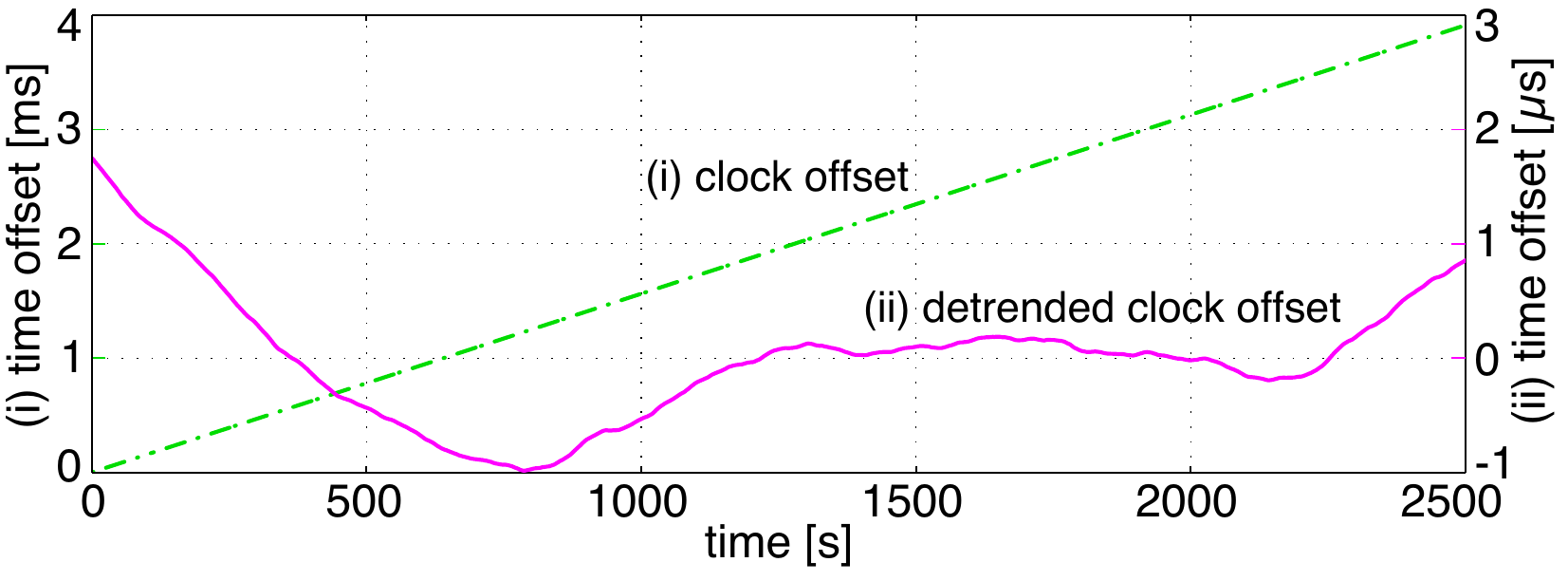}
\caption{Recorded (i) relative clock offset and (ii) detrended clock offset (clock noise), obtained from the sideband-carrier beat note phase measurement.}
\label{fig:clocks} 
\end{figure}

All data was recorded and processed offline, as will be the case for LISA. For ease of analysis, the phasemeters were started with a synchronization pulse. After the start, each clock ran independently. The time evolution of the clocks was measured from the sideband-carrier heterodyne phase, shown in Fig.\,\ref{fig:clocks}. Trace\,\ref{fig:clocks}i shows that the clock offset grows to 4\,ms over 2500\,s (1.6\,ppm frequency difference). The detrended clock offset, Trace\,\ref{fig:clocks}ii, shows the relative clock noise.

\begin{figure}[htb]
\includegraphics[width=8cm]{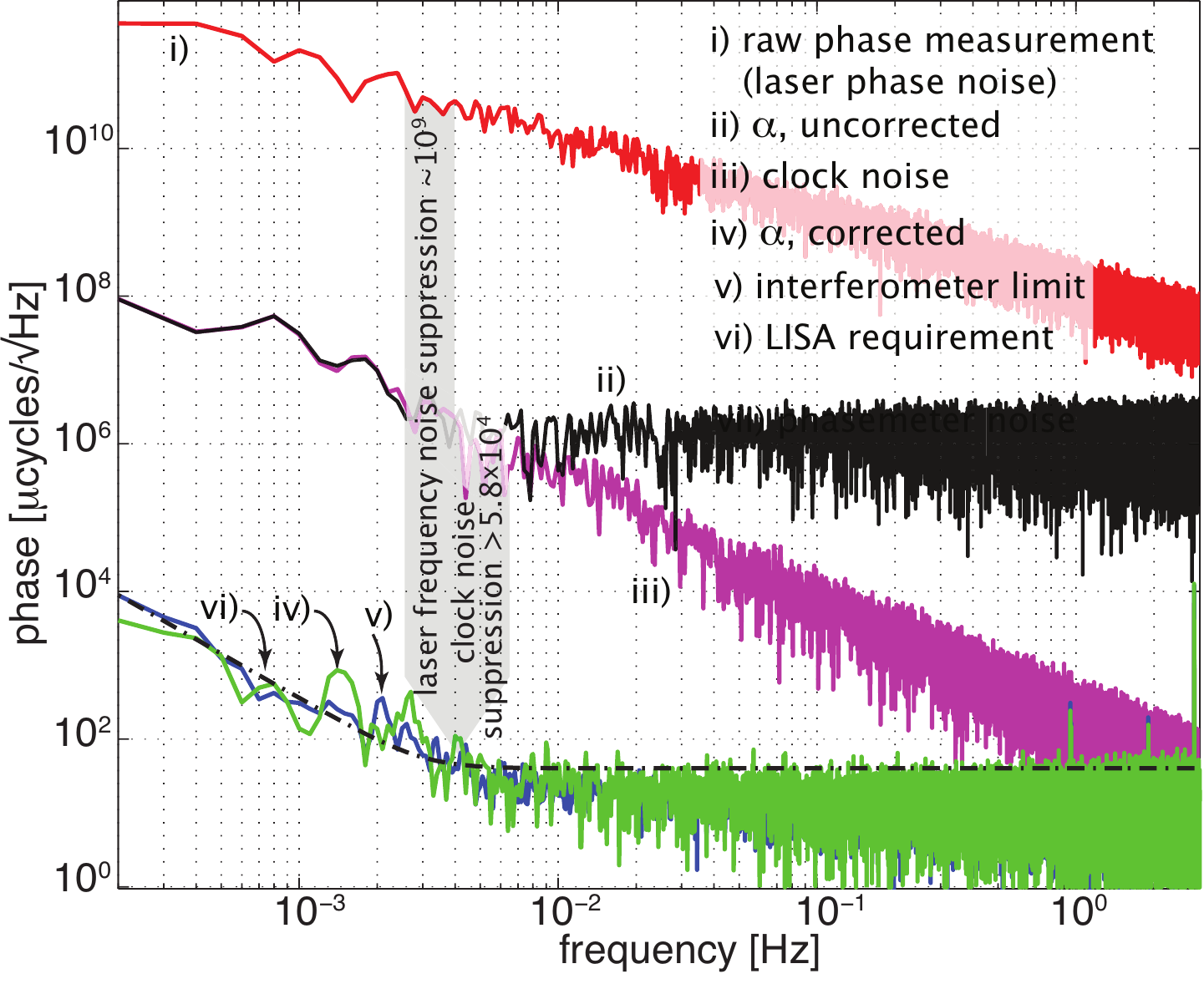}
\caption{Displacement measurements of the LISA interferometry test bed, showing: i) injected laser phase noise; iv) interpolated and clock 
noise corrected Sagnac TDI variable, demonstrating phase noise cancelation by $\sim$9 orders-of-magnitude, down to the interferometer 
noise floor, v). Note: 1$\mu$cycle $\simeq$ 1\,pm displacement equivalent.}
\label{fig:results} 
\end{figure}

Fig.\,\ref{fig:results} shows the experimental results, presented as root power spectral densities over the frequency range 0.2\,mHz to 3\,Hz. Trace \ref{fig:results}i shows the signal measured at photodiode $\beta_1$; the phase spectrum of the 800\,Hz/$\sqrt{\rm Hz}$ noise injected at the error point of the phase locking loop derived from $\beta_1$. This level of noise is present on all detectors and phasemeters, except $\beta_2$ which is used as the phase locking signal for laser $p_{32}$ to $p_{12}$.

If $\alpha$ is formed without correcting for either clock offset or clock noise, both laser and clock fluctuations couple into the measurement (as in Eq.\,\ref{eq:AlphaClocks}), shown in Trace\,\ref{fig:results}ii. To remove laser frequency noise with unsynchronized clocks, recorded signals from one bench were resampled by interpolation, synchronizing sampling times of the two benches' measurements (we assume negligible arm length, $D_{ij}\rightarrow1$). Using interpolation algorithms from\,\cite{Shaddock04}, we interpolated Bench 2 signals (the correction is symmetric, either bench could have been chosen):
\begin{eqnarray}
\alpha_{int}(t_1) &=&D_\Delta(\phi_{12}^c(t_2)-\phi_{32}^c(t_2)-2\beta_2^c(t_2))\nonumber \\ &&+
\phi_{31}^c(t_1)-\phi_{21}^c(t_1)-2\beta_1^c(t_1),\label{eq:alphaInt}\end{eqnarray}
Here we have defined the difference of the time measured by the two clocks, $\Delta t (t_1)= t_2-t_1$ and the delay operator, $D_{\Delta}
a(t) =  a(t+\Delta t(t))$. This removes laser frequency noise from $\alpha$ as shown in Trace\,\ref{fig:results}iii. However, clock noise is still present and now dominates the spectrum. For spacecraft 1, for example:
\begin{eqnarray}
\phi_{c1}(t_1)&=&\phi_{sb1}(t_1) - \phi_{21}^c(t_1)\nonumber \\& =& \left(q_2(t_1)-q_1(t_1)\right)f_{12}(t_1),
\end{eqnarray}
where $f_{12}(t_1)$ is the clock noise sideband microwave frequency. The clock noise and frequency noise free TDI combination is then:
\begin{eqnarray}
\alpha(t_1) = \alpha_{int}(t_1)-\frac{\phi_{c1}(t_1)\nu_{TB}(t_1)}{f_{12}(t_1)},\label{eq:alphaIntCNC}
\end{eqnarray}
where  $\nu_{TB}(t_1)=\nu_{12}(t_1)+\nu_{21}(t_1)-\nu_{31}(t_1)-\nu_{32}(t_1)$. This final TDI output is Trace\,\ref{fig:results}iv, reaching a displacement sensitivity similar to the total interferometry budget of LISA, Trace\,\ref{fig:results}vi. The final TDI output matches the measured displacement limit of the interferometer, Trace\,\ref{fig:results}v; the level with synchronized (phase locked) clocks and no noise injected into the laser phase locking.

The final TDI output shows laser frequency noise was suppressed by approximately $10^9$ at 3\,mHz, relative to the noise level in the individual phase measurements. LISA requires a laser frequency noise suppression factor of $3\times10^{7}\,{\rm Hz}/f$\,\cite{FCST}. The clock noise was suppressed by $6\times10^4$ at 3\,mHz, compared to the required suppression factor of 10-1000 (depending on the clock flown).

The original planning documents for LISA\,\cite{Bender} recognized the problem of suppressing phase noise from lasers and clocks. However, the proposed techniques were impractical and untested. In NASA's 2004 LISA Technology Development Plan, cancellation of laser noise and clock noise were rated 1st and 2nd on the project's 69 item list of interferometry technology risks. It would take the dual breakthroughs of post-processed TDI and high dynamic range phasemeters to realize a design that could be implemented. Work presented here tests these techniques in a configuration similar to LISA, including most of the significant noise effects. The results provide validation of the LISA phasemeter, the essential features of the LISA optical design, and evidence that LISA performance will not be limited by technical phase noise.

TDI represents a paradigm shift in the way interferometric measurements are performed. It is a move away from the conventional approach of stabilizing prior to measuring. Instead TDI allows measurements to be made using a noisy light source and relies on common-mode rejection of the noise by post-processing. By shifting the burden of noise rejection from hardware to signal processing, TDI has the potential to drastically simplify many interferometric measurement systems.

This research was performed at the Jet Propulsion Laboratory (JPL), California Institute of Technology (CIT), under contract with the National 
Aeronautics and Space Administration; in part, supported by appointments to the NASA Postdoctoral Program at JPL, administered by Oak Ridge Associated Universities via NASA contract. The authors acknowledge Peter Halverson, 
Akiko Hirai, Martin Regehr and Andreas Kuhnert for contributions. Copyright 2010 CIT. Government sponsorship acknowledged.


\begin{thebibliography}{9}

\bibitem{Bender} P.L. Bender, K. Danzmann and the LISA Study Team, Doc., lisa.gsfc.nasa.gov/Documentation/ppa2.08.pdf, MPQ {\bf 233}, (1998).

\bibitem{LISASC2007} J. Baker \textit{et al.} LISA Mission Science Office, LISA-LIST-RP-\textbf{436}, (2007).

\bibitem{Tinto99} M. Tinto and J.W. Armstrong, Phys. Rev. D \textbf{59}, 102003 (1999).

\bibitem{LISApm} D.A. Shaddock, et al., \textit{6th International LISA Symposium}, AIP Conference Series,  \textbf{873}, 654-660, (2006).

\bibitem{Dhurandhar} S.V. Dhurandhar, K.R. Nayak and J.-Y. Vinet, Phys. Rev. D \textbf{65}, 102002, (2002).

\bibitem{Shaddock03} D.A. Shaddock, et al., Phys. Rev. D \textbf{68}, 061303(R), (2003).

\bibitem{TintoPRD04} M. Tinto, F. B. Estabrook and J.W. Armstrong, Phys. Rev. D \textbf{69}, 082001, (2004).

\bibitem{Shaddock042} D.A. Shaddock, Phys. Rev. D \textbf{69}, 022001, (2004).

\bibitem{FCST} LISA Frequency control study team, {\it LISA Frequency Control White Paper}, ESA document LISA-JPL-TN-{\bf 823} (unpublished), (2009).

\bibitem{Shaddock04} D.A. Shaddock, et al., \prd \textbf{70}, 081101(R), (2004).

\bibitem{Hellings} R. Hellings, et al., Opt. Comm. {\bf 124}, 313, (1996).

\bibitem{Klipstein06} W. Klipstein, et al., 6th International LISA Symposium, AIP, (2006).

\bibitem{Shaddock03spie} D.A. Shaddock, B.C. Young and A. Abramovici, Proc. of SPIE, {\bf 4856}, 78, (2003).

\end{thebibliography}

\end{document}